\documentstyle[12pt,epsf]{article}
\begin{document}

%\pagestyle{empty}

%\begin{flushright}
%{\footnotesize Proc. 7th Int. Conf. on Atomically Controlled Surfaces, 
%Interfaces and Nanostructures, Nara, Nov. 2003}
%\end{flushright}

\begin{center}
%{\large
{\bf Design of electron correlation effects in 
interfaces and nanostructures}\par
\vspace*{0.3cm}
Hideo Aoki\par
%}
\vspace*{0.3cm}
{\it
Department of Physics, University of Tokyo, 
Hongo, Tokyo 113-0033, Japan
}
\end{center}
\ \\
Abstract 
We propose that one of the best grounds for 
the materials design from the viewpoint of 
{\it electron correlation} such as ferromagnetism, superconductivity 
is the atomically controlled nanostructures and heterointerfaces, 
as theoretically demonstrated here from three examples 
with first-principles calculations:  
(i) Band ferromagnetism in a purely organic polymer of 
five-membered rings, where the flat-band ferromagnetism 
due to the electron-electron repulsion is proposed.  
(ii) Metal-induced gap states (MIGS) of about one atomic monolayer thick 
at insulator/metal heterointerfaces, 
recently detected experimentally, for which 
an exciton-mechanism superconductivity is considered. 
(iii) Alkali-metal doped zeolite, 
a class of nanostructured host-guest systems, 
where ferromagnetism has been experimentally 
discovered, for which a picture of the 
``supercrystal" composed of ``superatoms" is proposed 
and Mott-insulator properties are considered. 
These indicate that design of electron correlation 
is indeed a promising avenue for nanostructures and heterointerfaces. 

%%% Introduction %%%%%%%%%%%%%%%%%%%%%%%%%%%%%%%%%
\section{Introduction}

There is a growing realisation that 
the concept of materials design is becoming 
realistic, which is due to advances in fabricating 
tailor-made materials and in performing 
first-principles calculations.  
Now, one decisive direction in the condensed-matter physics 
in these decades is the physics of electron correlation, i.e., 
the high-Tc superconductivity as kicked off by the cuprates, 
ferromagnetism, Mott's metal-insulator transition, etc.  
So we can ask a question: 
can we envisage {\it materials design from the viewpoint of 
electron correlation}\cite{fermiology}?  
Here we wish to advocate an idea that atomically controlled 
heterointerfaces and nanostructures, with their higher controllability 
than in the bulk, should be one of the best grounds to be explored.   
Specifically we present here proposals for one-, two- and 
three-dimensional systems, by combining the first-principles 
calculation with many-body studies: 
  
(1D) We propose ferromagnetism in purely organic 
polymers for the first time, 
where we show, with a computer-aided design, that the ferromagnetism 
due to the electron-electron repulsion 
in systems having flat one-electron bands should be realised 
in a chain of five-membered rings\cite{polyazolePRL,suwacrystal}. 

(2D) Second, we consider the metal-induced gap state (MIGS) 
at insulator/metal interfaces, which has been 
recently detected experimentally in LiCl/Cu,Ag for the first 
time\cite{kiguchiPRL}.  Our first-principles 
calculation\cite{kiguchiPRL,LiI} shows, despite the 
conventional wisdom, that the MIGS of 
about one atomic monolayer thick indeed exist.  
In the MIGS at insulator/metal interfaces the coupling of the 
carrier to excitons should be strong, 
which we envisage to work favourably for the exciton-mechanism 
superconductivity. 

(3D) Third, we take the alkali-metal doped zeolite, 
a class of nanostructured 
host-guest systems, where ferromagnetism has been experimentally 
discovered.  We propose to regard the system as a 
``supercrystal" composed of ``superatoms" (i.e., clusters of 
loaded atoms), where the first-principles 
band structure is so surprisingly simple as to allow such 
a picture\cite{zeolite}.  Mott-insulator properties 
can be understood as an effect of the electron correlation 
in the supercrystal.

%%%  polyazole %%%%%%%%%%%%%%%%%%%%%%%%%%%%%%%%%
\section{Band ferromagnetism in an organic polymer}

\begin{figure}
\begin{center}
\leavevmode\epsfysize=75mm \epsfbox{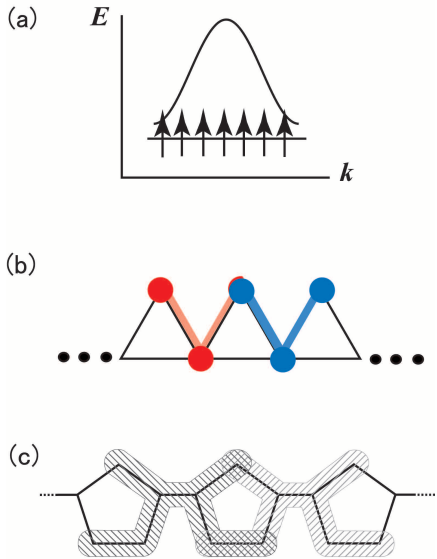}
\caption{(a) Flat-band ferromagnetism is schematically 
shown.  
(b) A chain of triangles, with two 
``Wannier" orbits that satisfy the connectivity condition indicated.  
(c) A chain of five-membered rings. 
}
\label{FBmodel}
\end{center}
\end{figure}

{\it Introduction} --- 
The discovery of conducting organic polymers\cite{Shirakawa74}
has kicked off intensive studies, including optoelectronics\cite{Friend99}. 
In the viewpoint of 
the electron correlation, design of (purely-)organic\cite{Allemand91}, 
ferromagnetism is a challenging target, since, 
usually, ferromagnetism is specific to d or f electron systems. 
A theoretical proposal was made by Shima and the 
present author\cite{Shima93} for ``superhoneycomb" 
networks of $\pi$-electrons, 
but this has to do with ferr{\it i}magnetism\cite{Lieb}.  
Here we propose a novel possibility
of a {\it band ferromagnetism} in an 
organic polymer, which is a realisation of 
the flat-band ferromagnetism (Fig.\ref{FBmodel}(a))
due to Mielke-Tasaki's mechanism.\cite{Tasaki-review}  

{\it Why the flat-band ferromagnetism?} --- 
We should first mention why we have to evoke a somewhat esoteric 
flat-band ferromagnetism.  While the problem of ferromagnetism in 
repulsively interacting electron systems has 
a long history dating back to Gutzwiller, Kanamori and Hubbard 
in the 1960's, we are still some way from a full understanding.  
In this context the flat-band ferrimagnetism 
proposed by Lieb\cite{Lieb} and the subsequent flat-band ferromagnetism 
by Mielke and by Tasaki\cite{Tasaki-review} are remarkable 
in that the magnetic ground state is proved rigorously for the 
repulsive Hubbard model when the one-electron band structure 
contains a flat band.  The flat band considered here is not the 
usual narrow band limit (with a set of disjointed basis functions), 
but those satisfying a special property 
called the local connectivity condition. There, 
adjacent ``Wannier" orbitals overlap (Fig.\ref{FBmodel}(b))
no matter how they are chosen, which is 
why spins tend to align due to Pauli's principle.  
When the (half-filled) flat band lies at the 
bottom of the band structure, the ferromagnetism is guaranteed 
for arbitrary strengths ($0<U\leq \infty$) of the electron-electron 
repulsion, while the ferromagnetism is expected for $U<U_c$ 
when the flat band sits in between dispersive ones, 
as has been confirmed for a model atomic quantum wire.\cite{Arita98} 

%%%%%%%%%%%%%%%%%%
%%%%%%%%%%%%%%%%%%%%%%%%%%

The connectivity condition is usually difficult to 
realise, but relatively easy 
for one-dimensional chain of molecules\cite{Arita98}, which 
is why we have opted for organic polymers.  Specifically, 
chains of five-membered rings should be 
promising, because we find it empirically easier to realise 
the flat bands for odd-membered rings, where 
the frustration tends to suppress 
antiferromagnetism in favour of ferromagnetism.  The tight-binding model 
for the connected five-membered rings
has indeed a flat band in appropriate, realistic conditions, 
where the eigenstates on the flat band  satisfy 
the connectivity condition (Fig.\ref{FBmodel}(c)).  

\begin{figure}
\begin{center}
\leavevmode\epsfysize=110mm \epsfbox{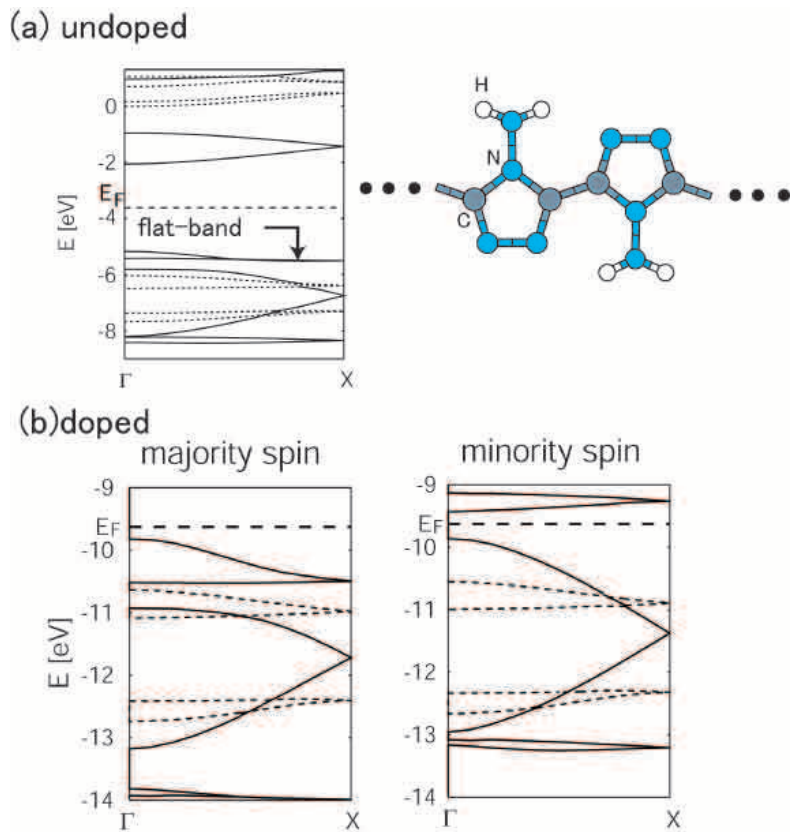}
\caption{The band structure of the undoped (a) 
and doped (b) PAT obtained by the GGA-SDFT.
The solid (dashed) lines represent 
bands with $\pi$ ($\sigma$) character.
The inset shows the (optimised) atomic configuration.  
}
\label{ldaband}
\end{center}
\end{figure}

{\it Search for the right polymer} --- 
So we start with a search for the case of flat bands by scanning 
various five-membered polymers, i.e., 
polypyrrole, polythiophene, etc.
The band structure is obtained with first principles calculations 
within the framework of the generalised gradient approximation 
based on the density functional theory 
(GGA-DFT).\cite{Perdew1996}
The atomic configuration is optimised to minimise the total energy 
with the conjugate gradient scheme\cite{Yamauchi1996}.
The flat band has turned out to be rather hard to realise even for 
five-membered chains (not a sufficient condition) even after 
various functional groups are attached to the 
rings, but we have succeeded in finding the right polymer, 
polyaminotriazole (PAT for short; see inset of Fig.\ref{ldaband}).  
Figure \ref{ldaband}(a) shows that the top valence band 
(with two branches corresponding to the band folding due to a dimerisation) 
has little dispersion ($\sim$ O(0.1eV)).

%%%%%%%%%%%%%%%%%%%%%%%%%%%
%%%%%%%%%%%%%%%%%%%%%%%%%%%%

We have then 
carried out the GGA calculation with the spin density functional theory 
(GGA-SDFT)\cite{Perdew1996} for the doped PAT with the half-filled flat band. 
The doping for the single chain is realised by increasing the number of 
electrons with a uniform positive background for charge neutrality.
The optimised state is polarised as shown in Fig.\ref{ldaband}(b)
with the splitting between the 
majority and minority-spin bands being $\sim 1$ eV, 
which is similar to the exchange splitting 
estimated in \cite{Okada00} for the atomic quantum wire.\cite{Arita98}
The ferromagnetic state has a total energy lower than that of 
the antiferromagnetic state by $\sim$50 meV.  
We have also shown that the 
Peierls distortion, to which 1D systems are prone, is 
negligibly small in the present system, which is 
because $\sigma$-electrons provide rigid enough backbone 
of the structure\cite{PIcomment}.   

\begin{figure}
\begin{center}
\leavevmode\epsfysize=50mm \epsfbox{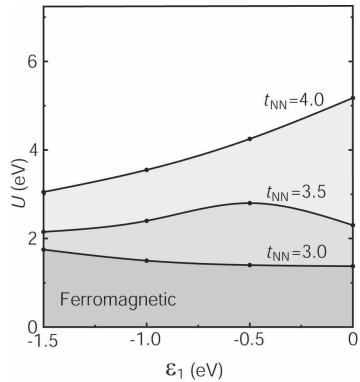}
\caption{Magnetic phase diagram for the Hubbard model 
exactly diagonalised for a finite (12-site) chain of 
five-membered rings against $U$ and 
$\epsilon_1$ (level of the bottom two nitrogen atoms 
relative to carbon) for various values of the N-N transfer 
$t_{\rm NN}$. 
}
\label{phase}
\end{center}
\end{figure}

{\it Electron correlation mechanism relevant} --- 
We can confirm that the ferromagnetism 
obtained in the band calculation is identified as the 
flat band ferromagnetism for the Hubbard model 
{\it a l{\'a}} Mielka-Tasaki.   To do so we have first mapped 
the $\pi$-electron system to a tight-binding model to examine 
whether the ground state is spin-polarised 
in the presence of the Hubbard repulsion, $U$.  
The phase diagram (Fig.\ref{phase}), drawn on a realistic 
parameter region, has a wide ferromagnetic region, to which 
PAT should belong (unless $U$ is unusually large ($> 4$ eV).  

%%%%%%%%%%%%%%%%%%%%%%%%%%%%
%%%%%%%%%%%%%%%%%%%%%%%%%%%%%%

\begin{figure}
\begin{center}
\leavevmode\epsfysize=120mm \epsfbox{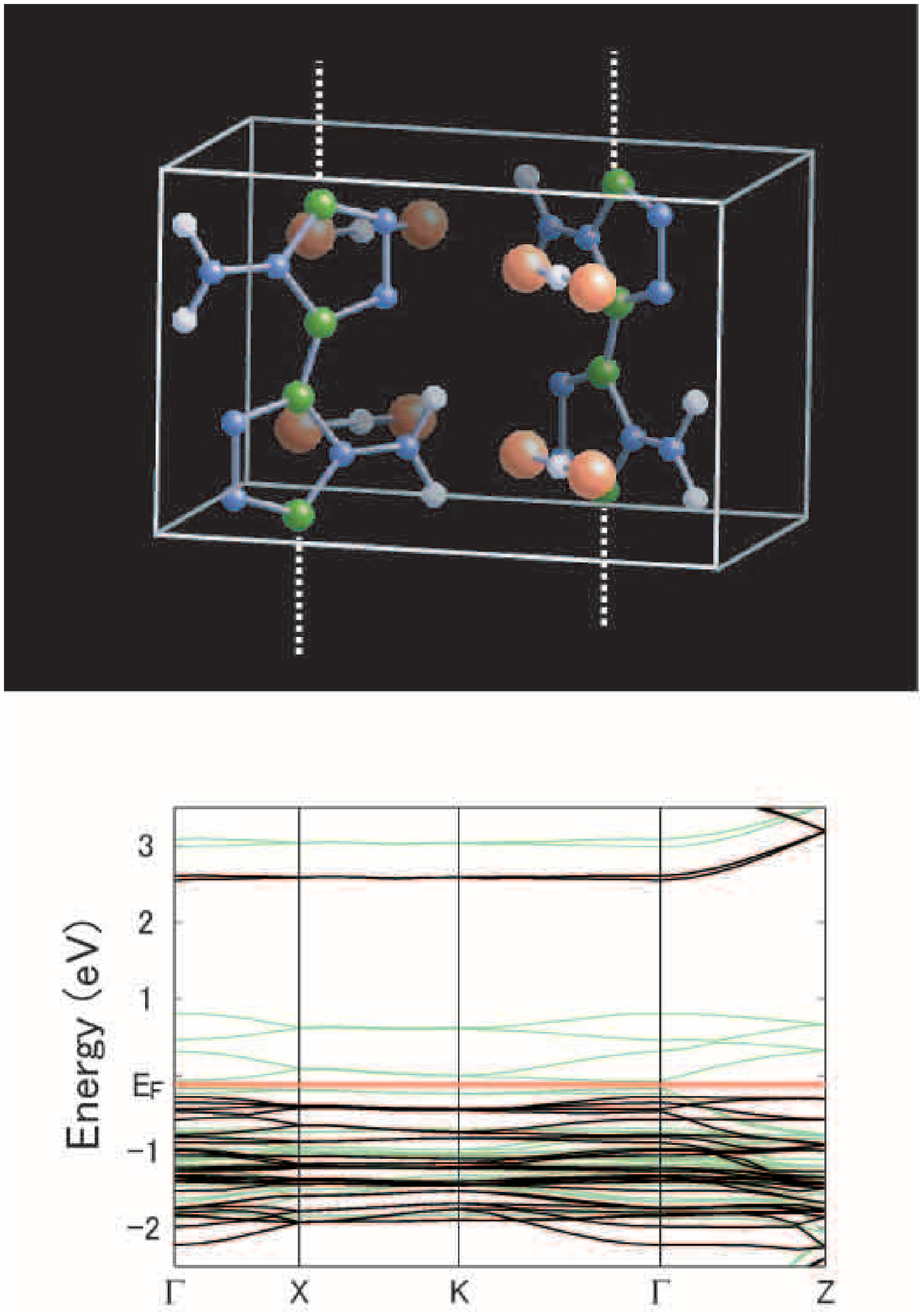}
\caption{
The optimised atomic configuration of the crystallised
polyaminotriazole doped with HF$_2$ (top), along with 
the band structure for the ferromagnetic solution with 
black (green) lines representing the 
majority (minority) spin.
%(c) The wave function (sum of the squared absolute values of 
%the top-four, majority-spin valence bands 
%at $\Gamma$) for the ferromagnetic polyaminotriazole crystal 
%doped with ClO$_4$.  Blue (red) contours represent 
%the amplitude around PAT (dopant), while 
%the green, white, blue, red, pink, orange balls represent the 
%position of the C, H, N, O, Cl and F atoms, respectively.  
}
\label{crystal}
\end{center}
\end{figure}

{\it Discussions} --- 
The flat-band ferromagnetism should work not only for 
long polymers but for 
oligomers as well, for which we should end 
up with high-spin state oligomers.  
Second, the one-electron dispersion does not have to be exactly 
flat\cite{Kusakabe,Penc}, nor does the band filling 
exactly half-filled\cite{Sakamoto} to 
realise the ferromagnetism.  
Also, electron-electron interactions can extend 
beyond  the on-site\cite{Shimoi}.
Even when a single chain becomes ferromagnetic, 
whether the magnetism remains in the bulk, i.e., in 
a three-dimensional crystal 
of polymers, is an important question.  We have 
studied this by means of the spin density functional 
calculation for the crystal\cite{suwacrystal} (Fig.~\ref{crystal}) 
to find that the ferromagnetism is robust against crystallisation, where 
the chemical dopant HF$_2$ puts the system 
close to the bulk ferromagnetism, so stronger anions 
such as BF$_4$ or PF$_6$ should be promising. 

%%%%%%%%%%%%%%%%%%%%%%%%%%%%%%%
%%%%%%%%%%%%%%%%%%%%%%%%%%%%%%%%

%%%% kiguchi %%%%%%%%%%%%%%%%%%%%%%%%%%%%%%%%

\section{Insulator/metal heterointerfaces --- 
metal-induced gap states}

{\it Why insulator/metal interfaces?} --- 
While there are mounting interests in the nature of 
heterointerfaces, insulator/metal interfaces are especially intriguing 
for their fascinating possibilities such as metal-insulator
transition\cite{jpc4,Anderson}, band gap narrowing \cite{prb64} and
superconductivity\cite{Ginzburgbook} as well as technological ones such as
catalysis, magnetic tunnelling junctions, etc. Despite the interest,
insulator/metal interfaces have not been studied satisfactorily, for good
reasons: well-defined interfaces are hard to prepare due to the
different nature of chemical bonds.    
  
A specific interest is the metal-induced gap states (MIGS), 
which were first introduced for
semiconductor/metal junctions in discussing the Schottky barrier at the
interface\cite{prl52}, and subsequently applied to insulator/metal
interfaces\cite{jpc8}. Roughly, MIGS are evanescent states 
of the metal electrons that (exponentially) penetrate into the insulating side
of the interface, with the penetration depth $\propto 1/E_G$.  
Recently, Muller {\it et al.} reported an observation of MIGS 
in MgO/Cu interfaces on fine particles \cite{prl80}. 
However, the interface studied there is polar (hence metallic presumably) and 
not well-defined atomically either, so there was 
some ambiguity. Kiguchi et al\cite{kiguchiPRL} have then made a clear-cut 
observation: they have employed 
atomically well-defined insulator/metal 
interfaces,\cite{prb63}
which are alkali halides grown epitaxially on metal substrates 
(LiCl on Cu(001) or Ag(001)). 
The near edge x-ray
 absorption fine structure (NEXAFS) 
has provided a clear evidence for the 
MIGS that are about one monolayer(ML) thick.  

\begin{figure}
\begin{center}
\leavevmode\epsfysize=145mm \epsfbox{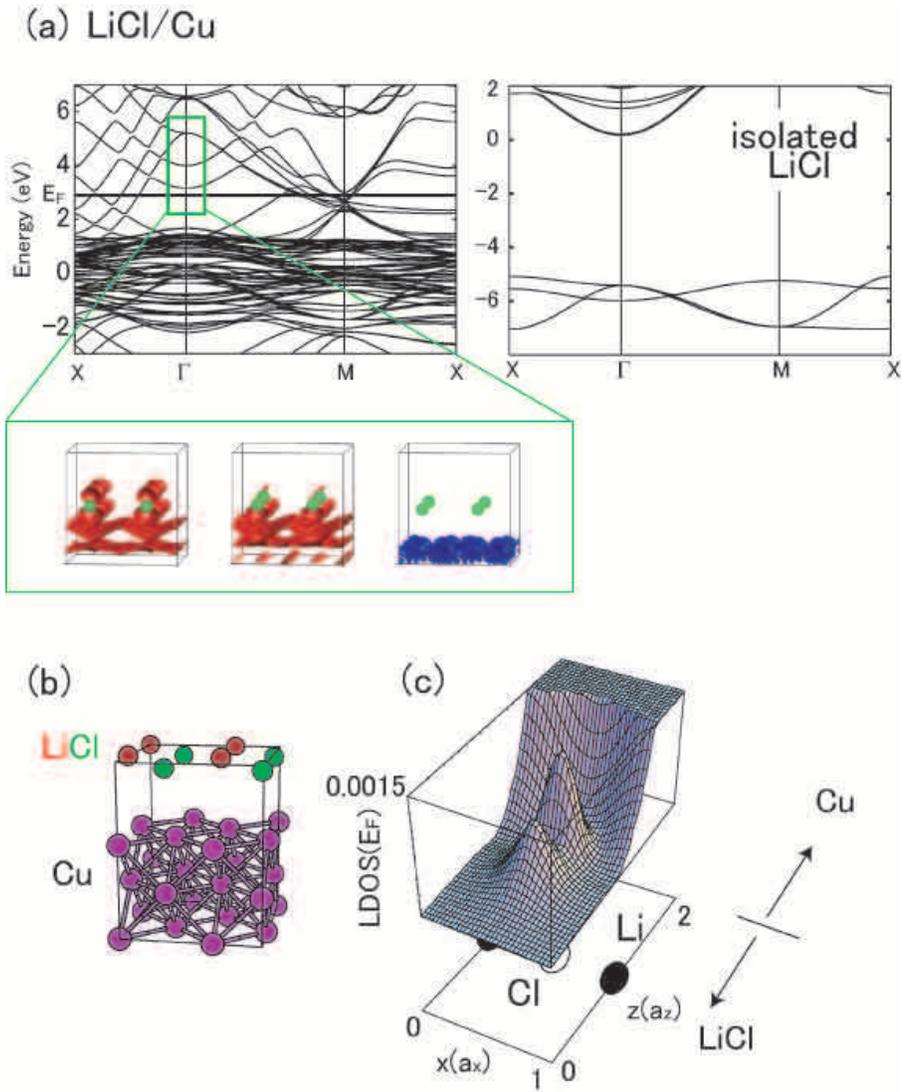}
\caption{(a) The band structure of 1 ML LiCl/Cu(001) as compared with 
that for an isolated 1 ML LiCl (right panel), along with 
the contours of the
absolute value of the LDF wavefunctions for the in-gap states
(MIGS in red).
(b) Atomic configuration.  (c) The charge distribution.}
\label{LiClCu}
\end{center}
\end{figure}

{\it Ab-initio calculation} --- 
So it is intriguing what an {\it ab-initio} calculation can tell us for the 
MIGS with a thickness $\sim$ 1 ML, since a simple model for the interface 
(e.g., tight-binding insulator/jellium) would predict a 
negligible ($\ll 1$ ML) penetration for the gap $E_G \sim O(10)$ eV.  
This is where the 
first-principles calculation based on the local density functional
 theory (LDF) comes in.  
Figure~\ref{LiClCu} (a) shows the band structure of 1 ML 
LiCl/Cu(001).  When the 1 ML
LiCl is put on Cu(001), new bands appear in the gap.  
In the LDF 
wavefunctions (Fig.~\ref{LiClCu}(c)) for the three bands just above E$_{F}$ 
at $\Gamma$ point, two in-gap states closest to $E_{F}$ indeed have, 
on top of the exponential decay, appreciable 
amplitudes on the interfacial Cl atoms 
with a p$_{z}$-like structure, in agreement 
 with the polarisation dependence in NEXAFS. 
MIGS with one ML thickness persist when we replace LiCl with LiF or LiI.
%(Fig.~\ref{LiI}).

It is well-known that the LDF generally underestimates the band gap, 
while GW approximation improves this.  
However, while the band-gap underestimation may be amended 
via the self-energy correction (i.e., without any
corrections for LDF wavefunctions), 
the shape of the wavefunctions should be reliable even in 
LDF, hence the present calculation for the local density of states 
is expected to be a good approximation.

%%%%%%%%%%%%%%%%%%%%%%%%
%%%%%%%%%%%%%%%%%%%%%%%%%%%

\begin{figure}
\begin{center}
\leavevmode\epsfysize=135mm \epsfbox{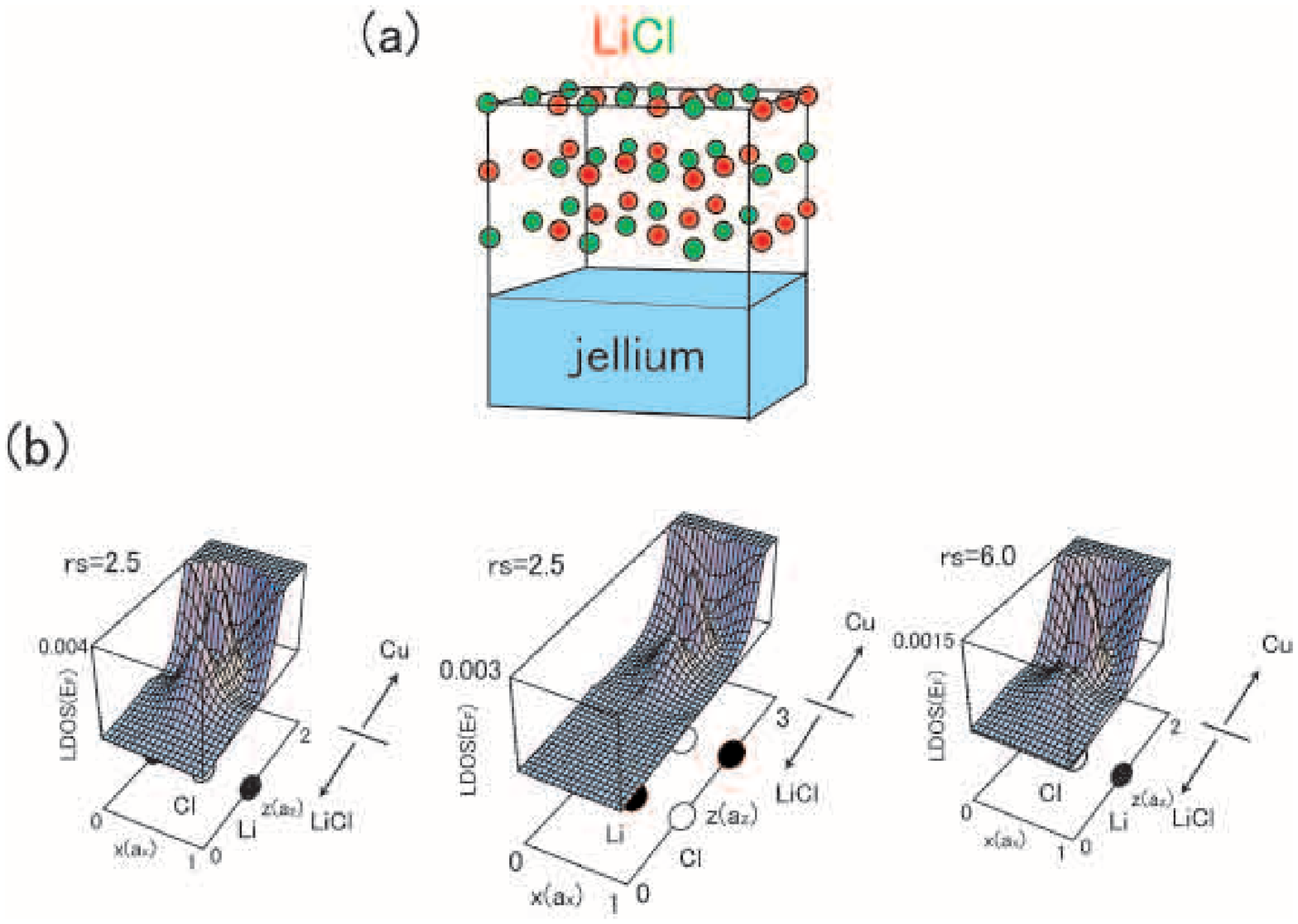}
\caption{Configuration (a) and 
the band structure(b) of 1ML(left) or 3ML(centre) 
LiCl/jellium with $r_{s}=2.5$.  The right panel is 
for $r_{s}=6.0$.  
}
\label{LiCljellium}
\end{center}
\end{figure}

We have also performed an LDF calculation for 3 ML 
LiCl put on a jellium 
to confirm the appearance of the MIGS is not specific to Cu 
(transition metal) substrate (Fig.~\ref{LiCljellium}).  
When we change the density of electrons in the metal 
by varying $r_s$ (the sole parameter characterising the jellium) 
the MIGS are found to be only 
weakly dependent on $r_s$ (Fig.~\ref{LiCljellium}). 
This accounts for the experimental result
that MIGS are observed for both Cu(with $r_s=2.7$) and Ag($r_s=3.0$) 
substrates.  
 
%%%%%%%%%%%%%%%%%%%%%%%%%%%%%
%%%%%%%%%%%%%%%%%%%%%%%%%%%%%%%%
 
{\it Electron correlation effects at the heterointerface} --- 
So what electron-correlation effects can we expect specifically in 
such insulator/metal interfaces?  A fascinating possibility should 
be the {\it exciton-mediated
superconductivity}. The exciton mechanism
has been originally proposed by Little\cite{Little64} for
quasi-one dimensional electrons 
to which polarisable molecules are attached, and 
subsequently two-dimensional version of this 
has been proposed for metal/semiconductor interfaces by 
Ginzburg\cite{Ginzburgbook,Ginzburg64,ABB}.   
However, a difficulty was subsequently 
pointed out by Inkson\cite{Inkson} that the presence of excitons
requires a wide-gap insulator, while this in turn prevents the metallic
carriers to penetrate into the insulator and makes 
the coupling of electrons to excitons too weak.  
 
\begin{figure}
\begin{center}
\leavevmode\epsfysize=45mm \epsfbox{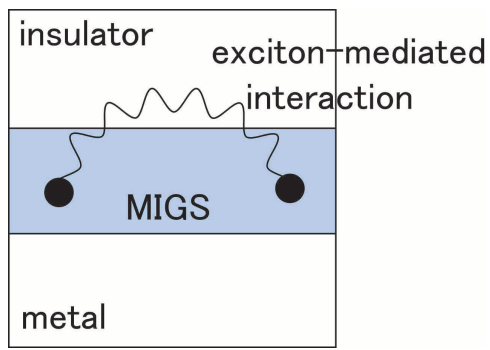}
\caption{Strong electron-exciton coupling schematically shown.
}
\label{exciton}
\end{center}
\end{figure}

In the present insulator/metal interfaces, by contrast, 
we do have a built-in coexistence (in real space) 
of carriers and excitations (associated with the 
wide band gap ($\simeq 9$ eV for LiCl) of the insulator), 
since the MIGS penetrate into the insulator side. So 
the carrier-exciton interaction should be strong (Fig.~\ref{exciton}).  
A theoretical estimate\cite{LiI} shows that the exciton contribution on top of 
the phonon mechanism can significantly enhance 
the superconductivity, where LiI on Al, for example, 
should make $T_c$ 1.6 times greater than that in the bulk Al. 
 
%%%%%%%%%%%%%%%%%%%%%%%%%%%%%%%
%%%%%%%%%%%%%%%%%%%%%%%%%%%%%
 
%%%%%%%%%%  Zeolite %%%%%%%%%%%%%%%%%%%%

\section{Electronic properties of alkali-metal loaded zeolites} 

{\it Why zeolites?} --- 
Materials design we have described so far is done 
by manipulating molecular or crystal structures.  
As an alternative way, we may envisage 
introducing guest atoms into 
host materials that are open-structured on nanometer scale.  
Doped zeolites are exactly such systems, and 
they are unique in that the host itself appears in a rich
variety of crystal structures, on top of rich possibilities for the
species and the number of guest atoms\cite{Bogomolv78}. 

\begin{figure}
\begin{center}
\leavevmode\epsfysize=60mm \epsfbox{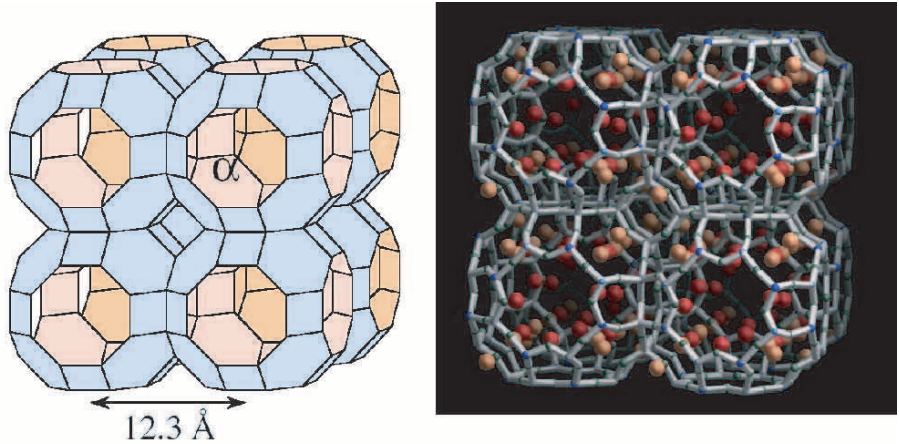}
\caption{
Left: Geometry of the LTA structure. Right: Unit cell 
of the zeolite LTA (undoped) with four $\alpha$ cages visible 
(dark blue: Si, light blue: Al, dark green: oxygen, orange: 
K on face centres of $\alpha$ cage, red: the other K). 
}
\label{LTA}
\end{center}
\end{figure}

The nanometer scale implies that the Coulomb interaction energy, 
which only decreases inversely with the size, can remain
large ($\sim$ eV) for the cluster doped in the cage.  
Indeed, experimental results by Nozue {\em et al} 
have established that some
zeolites (Fig.~\ref{LTA}) loaded with $\simeq$ five potassium atoms 
per cage are ferromagnetic for
$T<8$K\cite{Nozue92}, despite all the ingredients 
being non-magnetic elements\cite{commentmag}. 

%%%%%%%%%%%%%%%%%%%%%%%%%%%%
%%%%%%%%%%%%%%%%%%%%%%%%%

{\it Formulation and the result: an array of ``superatoms"} --- 
First question we want to ask is: can such host-guest systems 
(where the number of atoms in a unit cell is 
as large as 84) have simple electronic structures?   
So we start with the undoped aluminosilicate zeolite 
having a simple-cubic array of cages (Fig.~\ref{LTA}), where each cage
(called $\alpha$) is an Archimedes polyhedron. The region
surrounded by eight $\alpha$ cages forms another cage called
$\beta$. The material used in most experiments\cite{Nozue92} is
K(potassium)-form 
zeolite A (abbreviated as LTA) with a chemical formula 
K$_{12}$Al$_{12}$Si$_{12}$O$_{48}$.  
The crystal structure obtained from a recent 
neutron powder diffraction study by Ikeda {\it et al}\cite{Ikeda00} 
is adopted.

We have then performed an LDF calculation with the all-electron
full-potential linear muffin-tin orbitals (FP-LMTO)\cite{Mark}.  
In fact ours is the first reliable band structure calculation for 
zeolites\cite{Sun98}.  The result for the band structure 
for the undoped LTA shows that the undoped zeolite 
has the conduction bands 
whose wavefunctions primarily reside within the $\alpha$ cage 
or within $\beta$. 
The dispersion of the bands around the energy gap 
can be excellently fitted by the dispersion
of tight-binding bands on the simple cubic lattice, 
and we can identify the band as tight-binding bands 
of $s$- or $p$-like orbitals. 

\begin{figure}
\begin{center}
\leavevmode\epsfysize=98mm \epsfbox{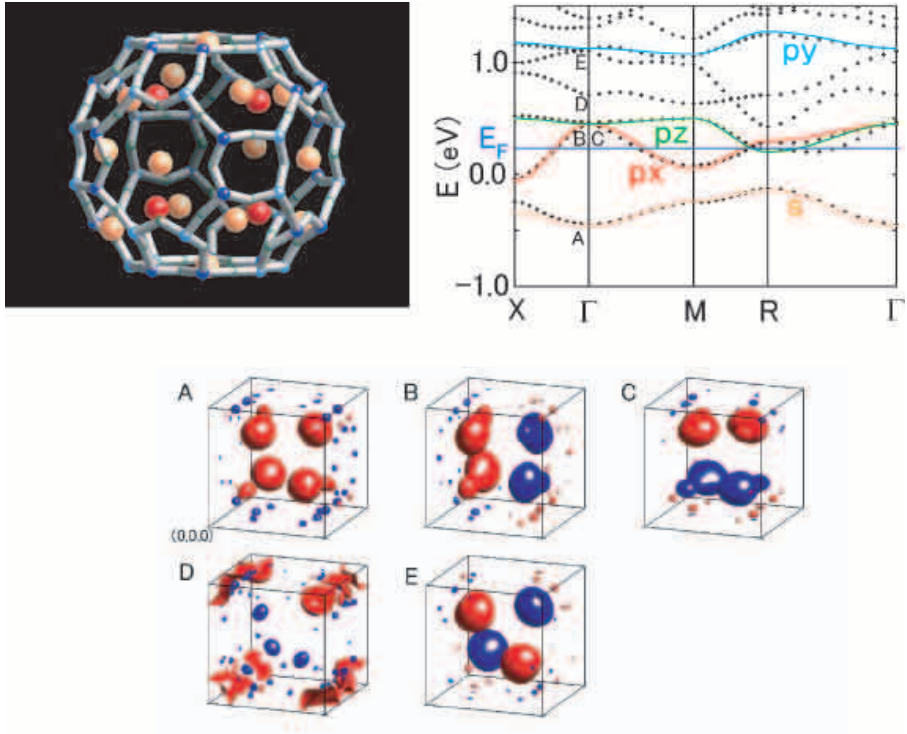}
\caption{
Band structure of a K-doped zeolite K$_3$LTA. Coloured curves are a 
tight-binding fit. Top left inset depicts the atomic configuration, 
where four red K's (three doped ones + the red one 
in Fig.\ref{LTA})
form a cluster in the $\alpha$ cage. Bottom panels are wave functions at 
$\Gamma$ in the bands A-E as labelled in the band structure.
}
\label{K3LTA}
\end{center}
\end{figure}

%{\it Doped zeolites} --- 
When three K atoms per unit cell are doped (K$_3$LTA, 
which belongs experimentally to the magnetic regime), 
three bands around $E_F$ can be identified as $p_x$, $p_y$, and $p_z$
bands, respectively, as shown in Fig.\ref{K3LTA}.  The fit of the dispersions to
the tight-binding model is again excellent, where 
the hopping integrals are almost an order of magnitude greater than 
those for K$_1$LTA, as expected from the larger cluster size.  
In passing we note the doped zeolites are 
rather distinct from the doped solid fullerene: The
latter, while a nanostructured crystal as well, 
has the relevant
orbitals that are basically LUMO/HOMO of the cage (buckyball) 
rather than those of dopants even after 
doped with alkali metals\cite{saitooshiyama}.  

%%%%%%%%%%%%%%%%%%%%%%%%%%%%%%%
%%%%%%%%%%%%%%%%%%%%%%%%%%%%%%

The good fit to a simple 
tight-binding model is highly nontrivial, since the ionic host, 
(K$^+)_{12}$(Al$^{3+})_{12}$(Si$^{4+})_{12}$O$^{2-})_{48}$, 
should possess a wildly varying potential well.  
Chemically, this should be because 
the cage has a low electron affinity so that the
electrons stay well away from the wall, as confirmed here. 
The surprisingly simple band structure around $E_F$ 
enables us to regard each cluster in the cage a 
``superatom" with well-defined $s$ and $p$ orbits, 
and regard the whole system as 
a ``supercrystal" (i.e., an array of superatoms).  
The result explains why the experimentally obtained 
optical spectrum can be assignment to those between
$s$ and $p$ orbitals in a simple well\cite{Kodaira93}.  

\begin{figure}
\begin{center}
\leavevmode\epsfysize=50mm \epsfbox{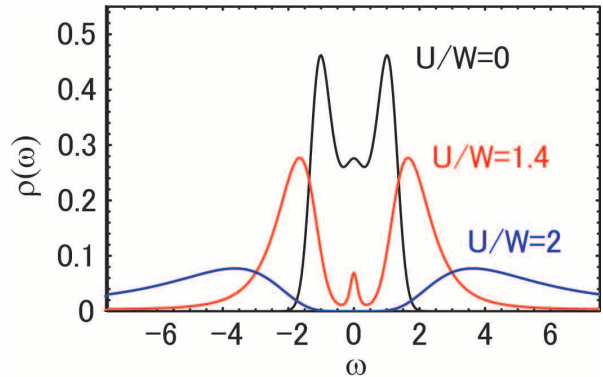}
\caption{
Spectral function, $\rho(\omega)$, for various values of $U/W$ 
obtained with the dynamical mean-field 
theory for the Hubbard model on 
an anisotropic cubic lattice ($t_x : t_y : t_z=$5:1:1).  
}
\label{Mott}
\end{center}
\end{figure}

{\it Electron correlation properties} --- 
Now we come to the question of whether 
the system is strongly correlated.  
The largest Coulomb matrix element 
is the intra-orbital Coulomb interaction $U$, which 
is estimated here to be $U\simeq 4.5$ eV for the $s$ band in
K$_1$LTA, and $\simeq 4.0$ eV for the $p_x$ band in K$_3$LTA.
Given that $U/W \sim 10\gg 1$, where $W \sim 0.4$ eV is the band
width, we can expect that these materials are Mott insulators.
However, since the relevant $p$ bands are 
significantly anisotropic, we have to be careful in estimating 
the critical $U_c$.  Here we have 
employed the dynamical mean-field theory\cite{Georges96} 
with the maximum 
entropy method\cite{Jarrell96} to estimate the transition point 
for a typically anisotropic 
($t_x : t_y : t_z=$5:1:1) single-band Hubbard model.  
The result shows that the system becomes a Mott insulator 
(as indicated by a gap in the spectral function) 
for $U/W> 2$ in the anisotropic case (Fig.\ref{Mott}).  
So we conclude that the K-doped zeolite is well on the
Mott-insulator side.  This resolves a puzzle that the band calculation
predicts that the system is a metal, while an infrared 
experiment indicates an insulator.\cite{Nakano99}
As for magnetic properties, 
the system is mapped to multi-band (triply-degenerate 
p-band) systems, which  are in general favourable for ferromagnetism 
since the inter-orbital kinetic-exchange coupling 
is ferromagnetic.  We 
can discuss magnetic properties in terms of the magnetic phase diagram 
obtained in the literature\cite{kusakabemulti,momoi}.

%%%%%%%%%%%%%%%%%%%%%%%%%%%%%
%%%%%%%%%%%%%%%%%%%%%%%%%%%%%%%

In the context of materials design, we expect that the electron correlation 
in zeolites can be controllable through
control of $U/W$.  For example, faujasite 
becomes metallic when metal-doped\cite{fau}, where
this form of zeolite has a significantly wider 
($7$\AA\ against $5$\AA\ for
LTA) window between the cages, 
which should result in a smaller $U/W$.

%%%%%%%%%%%%%%%%%%%%%%%%%%%%%%%%%%%%%%%%%
\section{Concluding remarks} 

The message obtained from the above examples is that 
the design of electron correlation 
can indeed be a promising avenue for atomically controlled 
nanostructures and heterointerfaces.  
There are a lot of theoretical works to be done, among which 
is the problem of how to estimate ``$U$".  Kusakabe, for instance, 
has shown that a kind of electron correlation Hamiltonian 
can be constructed by modifying the Kohn-Sham theory\cite{kabeU}.
Another way will be an elaboration of the LDF + dynamical mean-field 
theory.\cite{lda+dmft}  

The work described here are collaborations with 
R. Arita, Y. Suwa, K. Kuroki (ferromagnetic polymer), 
M. Kiguchi, Arita, G. Yoshikawa, Y. Tanida, 
M. Katayama, K. Saiki, A. Koma (heterointerface), 
Arita, T. Miyake, T. Kotani, 
M. van Schilfgaarde, T. Oka, Kuroki and 
Y. Nozue (zeolite).  
The works were supported in part by Grants-in-Aid for Scientific
Research and Special Coordination Funds 
from the Ministry of Education of Japan. 
Numerical calculations were performed with
SR8000 in ISSP, University of Tokyo, 
with the Tokyo Ab-initio Program Package for the GGA.

%%%%%%%%%%%% Figs polyazole

%%%%%%%%%%%%% Figs Kiguchi %%%%%%%%%%%%%%%%%%%%%%%%%%%%%%

%%%%%%%%%%%% Figs zeolite

%%%%%%%%%%%%%%%%%%%%%%%%%%%%%%%%%%%%%%%%


\begin{thebibliography}{99}
\bibitem{PIcomment} When the orbitals satisfy 
the connectivity condition, holes are doped into the system uniformly
so that there are no localised double bonds favouring 
the Peierls instability.  By contrast, polyacetylene, 
a conjugate polymer, has localised double bonds resonating with single bonds.
\end{thebibliography}
\end{document}